\documentclass[manuscript,screen,nonacm,oneside]{acmart}

\AtBeginDocument{%
  }

\setcopyright{none}

\newcommand{\toolname}{ExCritique}

\usepackage{array,longtable,booktabs}

\begin{document}

\title{ExCritique: An Interactive Visual Interface to Evaluate AI-Generated Explanations at Scale}


\author{Meheraj Hossain}
\affiliation{%
  \institution{University of Iowa}
  \city{Iowa City}
  \state{Iowa}
  \country{USA}}
\email{meheraj-hossain@uiowa.edu}

\author{Avinash Reddy Mudireddy}
\affiliation{%
  \institution{University of Iowa}
  \city{Iowa City}
  \state{Iowa}
  \country{USA}}
\email{avinashreddy-mudireddy@uiowa.edu}

\author{Naimul Hoque}
\affiliation{%
  \institution{University of Iowa}
  \city{Iowa City}
  \state{Iowa}
  \country{USA}}
\email{naimul-hoque@uiowa.edu}

\renewcommand{\shortauthors}{Hossain et al.}


\begin{abstract}
 Vision-language Models (VLMs) can generate natural language explanations for complex decision-making scenarios. However, these explanations can sometimes be misleading, sounding convincing while containing incorrect or unsupported information. Automated explanation-quality metrics help researchers and practitioners evaluate large collections of explanations, but their scores alone provide limited support for identifying failure cases and understanding the reliability of these metrics. This paper presents a case study on how interactive visualizations can help researchers and practitioners evaluate both AI-generated explanations and automated explanation-quality metrics across large datasets. Based on formative interviews with nine AI researchers and practitioners, we designed an interactive visualization tool named \toolname{}, situated in the task of Visual Question Answering (VQA). \toolname{} visualizes the distributions of multiple automated quality metrics, allows users to  explore thousands of data items, filter items and inspect the evidence behind individual scores. We compared \toolname{} with a baseline interface without interactive visualizations in a controlled study involving 16 AI researchers and practitioners. Our findings suggest that the interactive visualizations in \toolname{} helped users identify explanation-quality issues more efficiently and better understand automated quality metrics, supporting an effective way to measure the reliability of AI-generated explanations.
\end{abstract}

\ccsdesc[300]{Computing methodologies~Artificial intelligence}
\ccsdesc[500]{Human-centered computing~Visual analytics}

\keywords{Data Visualization, Explainable AI, HCI, Vision-language models, Explanation quality metrics}


\maketitle

\section{Introduction}
\label{sec:introduction}

Vision-language models (VLMs) can answer questions about images and provide natural-language explanations for their predictions. These explanations offer users a way to examine the evidence presented for an answer, but their fluency is not a guarantee of quality. A model may describe an image convincingly while overlooking a detail that determines the correct answer, or justify a prediction that the image does not support~\cite{kayser2021evil,he2026believing}. Moreover, agreement between an explanation and a prediction does not establish that the explanation reflects the model's internal decision process~\cite{parcalabescu2024measuring}. Evaluating VLMs therefore requires attention not only to the answers they produce, but also to what their explanations say and whether the available evidence supports those claims.

Recent work has developed automated quality metrics to make this assessment more systematic. Metrics such as FaithScore and VALOR-EVAL examine the visual grounding of generated content~\cite{jing2024faithscore,qiu2024valor}, while He et al. introduce \emph{Visual Fidelity} and \emph{Contrastiveness} to assess whether a VLM explanation is supported by the image and distinguishes its predicted answer from alternatives~\cite{he2026believing}. VLM-based judges can also evaluate outputs according to specified criteria~\cite{lee2024prometheusvision}. These approaches make it possible to assess large collections of outputs without reviewing every explanation manually. Yet the resulting scores are themselves estimates. The broader literature on LLM evaluation has documented how methodological choices and limitations can undermine the reliability of evaluation results~\cite{laskar2024survey}, and studies of model-based judges have exposed systematic evaluation biases~\cite{wang2024unfair,zheng2023judging}. For a practitioner using such scores, a low rating may point to a genuine explanation problem, but it may also reflect the judge's interpretation of the evidence or the assumptions built into the metric. The implications of these limitations extend beyond the accuracy of automated evaluation. He et al. ~\cite{he2026believing} show that presenting explanation-quality scores can influence how users judge the correctness of VLM predictions. However, the usefulness of such information depends on the reliability of the underlying evaluation. For researchers and practitioners, this raises the question of how they can determine whether the scores themselves reflect meaningful explanation-quality problems or limitations of the automated evaluators?

This creates a practical challenge that is not resolved by producing more scores. When reviewing hundreds or thousands of explanations, practitioners need to decide where to look, determine why an example received a particular rating, and distinguish recurring explanation failures from questionable automated judgments. A distribution may reveal that many explanations receive low Visual Fidelity scores, for example, but it does not show whether those explanations make unsupported visual claims or whether the verification procedure has misinterpreted them. Conversely, an individual explanation may receive high scores on one dimension while omitting information that is necessary for answering the question. Understanding these cases requires moving between dataset-level patterns and the image, question, prediction, explanation, and evidence underlying each assessment.

Interactive evaluation systems provide a promising approach for such investigations. Prior research demonstrates how interaction can help people investigate large collections of AI outputs rather than relying on isolated examples or summary statistics~\cite{kim2026evalet, huang2026vipera, yeh2025exploring}. For example, \emph{Vipera} uses visual guidance to support systematic auditing of text-to-image generation~\cite{huang2026vipera}. 
However, interactive workflows for jointly exploring VLM explanations and automated quality metrics remain unsolved.



We investigate this problem through \emph{\toolname{}}, an interactive visual interface for examining VLM-generated explanations at the dataset scale. Informed by formative interviews with nine AI/ML researchers and practitioners, we developed \toolname{} as a design probe to study how coordinated visual exploration can support explanation evaluation. We use multiple-choice visual question answering as an application setting in which explanations can be examined against images, questions, and alternative answers. The interface combines distributions of three complementary quality measures---Visual Fidelity, Contrastiveness, and our proposed \emph{Concept Attribution}, a measure of question-relevant visual-concept coverage---with a semantic map, interactive filtering, and detailed inspection of explanations and quality scores. Rather than reducing explanation quality to a single ranking, \toolname{} allows practitioners to investigate disagreements among metrics and examine the context and evidence of individual data items. The interface also provides a view for comparing evaluations across multiple models. 


We evaluated \toolname{} in a within-subject study with 16 participants who used both the proposed interface and a spreadsheet-style baseline to investigate potentially problematic explanations and automated scores. Participants identified more examples with potential issues when using \toolname{} and reported better understanding of explanation quality and metric performance. Their exploration patterns and interviews also revealed an important trade-off: coordinated views helped some participants investigate score relationships and locate unusual cases, whereas others preferred the baseline's simpler presentation for focused inspection. These findings inform the design of tools that support human scrutiny of both VLM explanations and the automated quality metrics used to evaluate them.

This paper contributes: (1) empirical findings about practitioners' needs and challenges when evaluating model-generated explanations with automated quality scores; (2) the design and implementation of \toolname{}, which connects dataset-level explanation-quality patterns to multimodal evidence and score-level inspection; and (3) a mixed-method evaluation of the opportunities and limitations of interactive visual exploration for examining VLM explanations and their automated quality scores.
\section{Related Work}
\label{sec:related-work}

We position \toolname{} at the intersection of automated explanation evaluation and interactive approaches for examining model behavior and evaluation results.

\subsection{Automated Evaluation of AI-Generated Explanations}
\label{sec:rw-explanation-evaluation}

Evaluating explanations is different than measuring quantitative performance measures such as accuracy. Research on natural-language rationales has examined properties such as evidence alignment, informativeness, and consistency between reasoning steps~\cite{deyoung2020eraser,chen2023rev,prasad2023receval}. Other methods assess generated text through its factual claims or correspondence with supporting sources~\cite{min2023factscore,zha2023alignscore}. The use of language models as evaluators has broadened the range of qualities that can be assessed automatically, including criteria specified through prompts or rubrics~\cite{liu2023geval,kim2024prometheus}. At the same time, automatic evaluation is not free from judgment errors. Laskar et al.'s systematic review identifies broader challenges to reliable and reproducible LLM evaluation~\cite{laskar2024survey}, while studies of LLM-based judges document biases and sensitivity to evaluation design~\cite{wang2024unfair,zheng2023judging}. Tests based on an explanation's consistency with model outputs must also be distinguished from tests of faithfulness to the model's internal reasoning~\cite{parcalabescu2024measuring}.

For VLMs, assessing explanations additionally requires checking what is actually present in an image. Benchmarks such as e-ViL connect visual inputs with natural-language explanations~\cite{kayser2021evil}; subsequent evaluation methods examine object hallucinations and finer-grained visual grounding~\cite{li2023pope,jing2024faithscore,qiu2024valor}. Prometheus-Vision develops a VLM-based evaluator for image-aware, user-defined criteria~\cite{lee2024prometheusvision}. Most directly related to our work, He et al. propose Visual Fidelity and Contrastiveness to assess whether an explanation is supported by the visual context and whether it favors the predicted answer over alternatives~\cite{he2026believing}. Their study also shows how displaying such quality scores can help users judge the reliability of model predictions in settings where the underlying image is not visible.

Together, these methods provide ways to generate quality assessments of explanations. We focus on the subsequent \emph{use} of these assessments by practitioners. A numerical rating does not necessarily indicate which visual evidence was considered, whether a crucial concept was omitted, or why different metrics disagree on the same example. \toolname{} addresses this gap by treating explanations, quality metrics, and the evidence used to obtain that metrics as related materials for human inspection.


\subsection{Interactive Evaluation and Auditing of AI Models}
\label{sec:rw-interactive-evaluation}

Interactive analysis has long been used to supplement aggregate model-performance measures with closer examination of examples and data subsets. Systems such as Errudite and the Language Interpretability Tool help practitioners inspect model behavior across cases~\cite{wu2019errudite,tenney2020lit}; FairVis, Domino, and SliceTeller support the discovery or comparison of subsets where model behavior differs~\cite{cabrera2019fairvis,eyuboglu2022domino,zhang2023sliceteller}. These approaches illustrate the importance of moving between an overview of a collection and the individual records responsible for a pattern. Their primary targets, however, are model predictions, test cases, or dataset slices rather than the quality of explanations and the trustworthiness of explanation-specific scores. Recent work extends interactive evaluation to generative models~\cite{jaunet2022visqa, huang2026vipera, arawjo2024chainforge, yeh2025exploring}. ChainForge supports comparisons across prompts and generated outputs~\cite{arawjo2024chainforge}, and LLM Comparator brings together overall results, flexible slicing, and side-by-side responses for interpreting automatic pairwise evaluations~\cite{kahng2025comparator}. Particularly close to our motivation, Evalet addresses the difficulty of validating LLM-as-a-judge outputs: it links judgments to functional fragments of generated text, enabling users to inspect and revise assessments at a more meaningful level than a holistic score~\cite{kim2026evalet}. Our work shares this concern for the interpretability of automated evaluations, but examines a different form of evidence. For VLM explanations, users must relate generated text not only to an evaluation criterion but also to the input image, visual facts, answer alternatives, and the intermediate results of multiple scoring procedures. Our interaction support focuses on visual exploration, which is also different than Evalet's text-based interactions.



\section{Formative Study}
\label{sec:formative-study}

We conducted formative interviews with nine AI/ML practitioners and researchers (P1-9) to understand how they evaluate AI- and VLM-generated explanations, the challenges they encounter, and their needs for this task.

\subsection{Participants and Study Procedure}


Five participants were research scientists at large technology companies, two were PhD students engaged in AI/ML development and evaluation, and two were professors actively overseeing AI/ML research projects. Overall, their experiences covered areas such as model development, AI-assisted coding, agentic AI systems, and model evaluation. Each participant was interviewed individually for approximately 30-45 minutes. 

The interviews consisted of three parts. First, we asked participants about their backgrounds, current work, and experiences with AI/ML. Second, we discussed situations in which they or their teams needed to evaluate explanations generated by LLMs or VLMs.  We asked about their evaluation processes, quality metrics, and the challenges they faced, particularly when evaluating explanations at the dataset scale. Three participants did not have to evaluate explanation quality directly; their feedback was based on experiences in relevant tasks like evaluating translations or summaries.
In the final part, we invited participants to brainstorm potential solutions and discuss the information, features, and interactions they would expect from an interactive explanation-evaluation tool.

\subsection{Findings}


\subsubsection{Differences in practice.}
All participants mentioned that automated judges and metrics are becoming a popular choice for measuring explanation quality and similar tasks (P1–P9). However, we observed differences in how industry and academic researchers approach automated evaluation. Industry participants reported that automated judges are almost always the default for evaluating open-text generation tasks, as their work demands quick turnaround and human annotations are often infeasible except for long-term research (P1–P5). In contrast, academic researchers were more open to collecting human annotations. P8, an Associate Professor, explained that they typically collect human annotations for a small subset of the data and compare them with automated evaluations across the larger dataset. Thus, while practices differ, both sets of users see automated metrics as the prevalent way to measure explanations quality.


\subsubsection{Quality metrics are task-dependent, but factualness is a common metric}
Participants emphasized that there is no single definition of a good explanation. As P2 explained, \emph{``The definition of goodness therefore depends on the task that we are trying to accomplish.''} However, five participants mentioned factualness as a useful objective metric to calculate. Other metrics mentioned by participants are relevance, logical consistency, clarity, and the presence of important contextual information. When asked about VLM-generated explanations, P4 suggested evaluating whether explanations include relevant visual objects and contextual information.




\subsubsection{Automated quality scores are difficult to interpret and verify.}

Although participants considered automated evaluation useful, they raised concerns about interpreting quality scores and determining whether those scores accurately reflect explanation quality. P6 stated, \emph{``The most important issue is making sense of the score,''} and added, \emph{``The system should also explain how it obtained that score.''} P5 expressed a related concern about existing evaluation metrics: \emph{``the metrics, sometimes they do not have fixed answers. It could be subjective and vague.''} Participants also emphasized the importance of human verification. P4 explained that \emph{``if the score is aligned with my judgment, then I trust it.''} These concerns highlight the difficulty of assessing the reliability of automated judgments based on numerical scores alone. 



\subsubsection{Dataset scale is a common problem}
Participants repeatedly mentioned that large datasets motivate the use of automated metrics. However, the scale also makes it difficult to examine individual examples and develop a broader understanding of explanation quality and the evaluation metrics. P2 highlighted this challenge, noting that \emph{``At the dataset scale, the main challenge is time,''} as reviewing each response individually becomes impractical. Describing their current workflow, P2 added, \emph{``In practice, we do not inspect every example. We usually sample examples, which is not very satisfactory.''} The quality metrics can guide sampling (e.g., filtering samples with low scores).   P1 noted that \emph{``The first thing I would want to see is the distribution of results,''} followed by \emph{``examples from that distribution.''} Similarly, P7 stated, \emph{``If I have one million explanations and one million probabilities, I want to visualize the distribution of those probabilities.''}






\section{\toolname{} System}
\label{sec:system}


The formative study highlighted the need for mechanisms that enable AI practitioners and researchers to efficiently understand the quality of AI-generated explanations and the automated metrics used to evaluate the explanations. Such mechanisms can help them identify faulty and misleading explanations and quality metrics, as well as develop strategies for improving both explanations and metrics. To address this need, we developed \toolname{}, a human-AI collaborative visual analytics system for critiquing explanations. This section describes overall design and technical details of \toolname{}.

\subsection{Design Goals}
We decided to develop \toolname{} as a \textit{technology probe}. According to Hutchinson et al.~\cite{10.1145/642611.642616}, ``Technology
probes are simple, flexible, and adaptable technologies with the
goal of understanding user needs and desires in a real-world set-
ting, field-testing the technology, and motivating the design of new
technologies.'' We decided to adapt this method since we wanted to understand the benefits and limitations of visual analytics for evaluating AI-generated explanations with automated quality metrics.
In the following, we describe design goals that drive the design of \toolname{}.

\textbf{DG1. Visual Question Answering (VQA) as the Application Domain.} Participants mentioned many scenarios and tasks where VLMs generate explanations. Supporting all scenarios and tasks is not feasible for a single technology probe. Thus, we decided to fix an application domain for \toolname{}. We chose visual question answering (VQA) as the application domain for the probe. VQA provides a structured setting in which models answer natural-language questions about images, often requiring both visual understanding and contextual reasoning~\cite{schwenk2022aokvqa,kayser2021evil}. This allows us to examine whether generated explanations contain relevant visual evidence, make unsupported claims, or omit important information. VQA is a widely used task in NLP and HCI research~\cite{bigham2010vizwiz, kayser2021evil}. Thus, the task will enable us to situate our work in the broader literature. 


\textbf{DG2. Show distribution for multiple quality metrics.} Participants mentioned that the critiquing task often starts with seeing the distribution of scores from automated quality metrics. Our tool should support that. In addition, the interview findings indicate that participants often examine multiple competing quality metrics. Our tools should compute and integrate multiple metrics in the interface. Shneiderman's \textit{Overview first, zoom and filter, then details-on-demand} principle can be applied here~\cite{shneiderman1996eyes}.

    
\textbf{DG3. Support flexible data slicing.} One of the challenges mentioned by participants was identifying data slices that require close verification. This is not always straightforward. Researchers and practitioners often look for unusual slices where quality metrics disagree in different combinations. Our tool should support flexible data slicing based on filtering the distributions of the quality metrics.


\textbf{DG4. Support for an individual data item.} The next step after data slicing is to investigate individual data points under the slice. For a coherent sensemaking, the system should present the input image, question, model answer, generated explanation, and corresponding quality scores all together for an individual item. Users should be able to understand what each score measures and examine the available evidence to assess whether the assigned scores are consistent with the explanation.

\textbf{DG5. Explain quality metrics.} Another key challenge mentioned by participants was interpreting the quality scores. Aggregate metric scores can hide how individual scores were calculated. For quick comprehension, our tool should visually explain the steps for calculating an individual score. Prior research has used visual forms to explain AI-based decision-making~\cite{wexler2020whatif, hohman2020summit}, which can be applied in the tool.


\subsection{Explanation-Quality Metrics}
\label{sec:quality-metrics}

To support the evaluation of VLM-generated explanations for VQA (DG1), \toolname{} incorporates three complementary quality metrics: \emph{Visual Fidelity}, \emph{Contrastiveness}, and \emph{Concept Attribution}. We adopt Visual Fidelity and Contrastiveness from He et al.~\cite{he2026believing} and introduce Concept Attribution to assess an additional aspect of explanation quality. Each metric produces a score between 0 and 1, with higher values indicating better performance on the corresponding dimension.

\textbf{Visual Fidelity.}
Visual Fidelity measures whether the visual claims made in an explanation are supported by the input image~\cite{he2026believing}. Following He et al., visual claims in the explanation are converted into verification questions. A verifier VLM answers these questions using the image, and the score is calculated as the proportion of questions answered affirmatively. A higher score indicates that more of the explanation's visual claims are supported by the image.

\textbf{Contrastiveness.}
Contrastiveness measures how strongly an explanation supports the model's predicted answer relative to alternative answers~\cite{he2026believing}. The predicted answer and other answer options are first masked in the explanation to reduce answer leakage. An entailment model then estimates how strongly the masked explanation supports a statement corresponding to each answer option. The score is computed by dividing the entailment probability of the predicted answer by the sum of entailment probabilities across all options. 


\textbf{Concept Attribution.}
Both Visual Fidelity and Contrastiveness are explanation-driven. They align the content of the explanation with the input image (explanation $\rightarrow$ input), but do not capture important input concepts that might be missing in the explanations (input $\rightarrow$ explanation). To address this issue, we introduce a new metric named concept attribution. The metric is motivated by prior work focusing on visual concepts as interpretable tokens of an image~\cite{hoque2022visual}.
Given an image $I$ and a question $Q$, we identify a set of relevant visual concepts $\mathcal{C}(I,Q)$. We then determine whether each concept is explicitly or implicitly mentioned in the explanation $E$. The Concept Attribution score is defined as

\begin{equation}
S_{\mathrm{CA}}(I,Q,E) =
\frac{\sum_{c \in \mathcal{C}(I,Q)} M(c,E)}
{|\mathcal{C}(I,Q)|},
\label{eq:concept-attribution}
\end{equation}

where $M(c,E)$ equals 1 if concept $c$ is mentioned in the explanation and 0 otherwise. A higher score indicates that the explanation covers a larger proportion of the relevant visual concepts.


\begin{figure}
    \centering
    \includegraphics[width=0.95\linewidth]{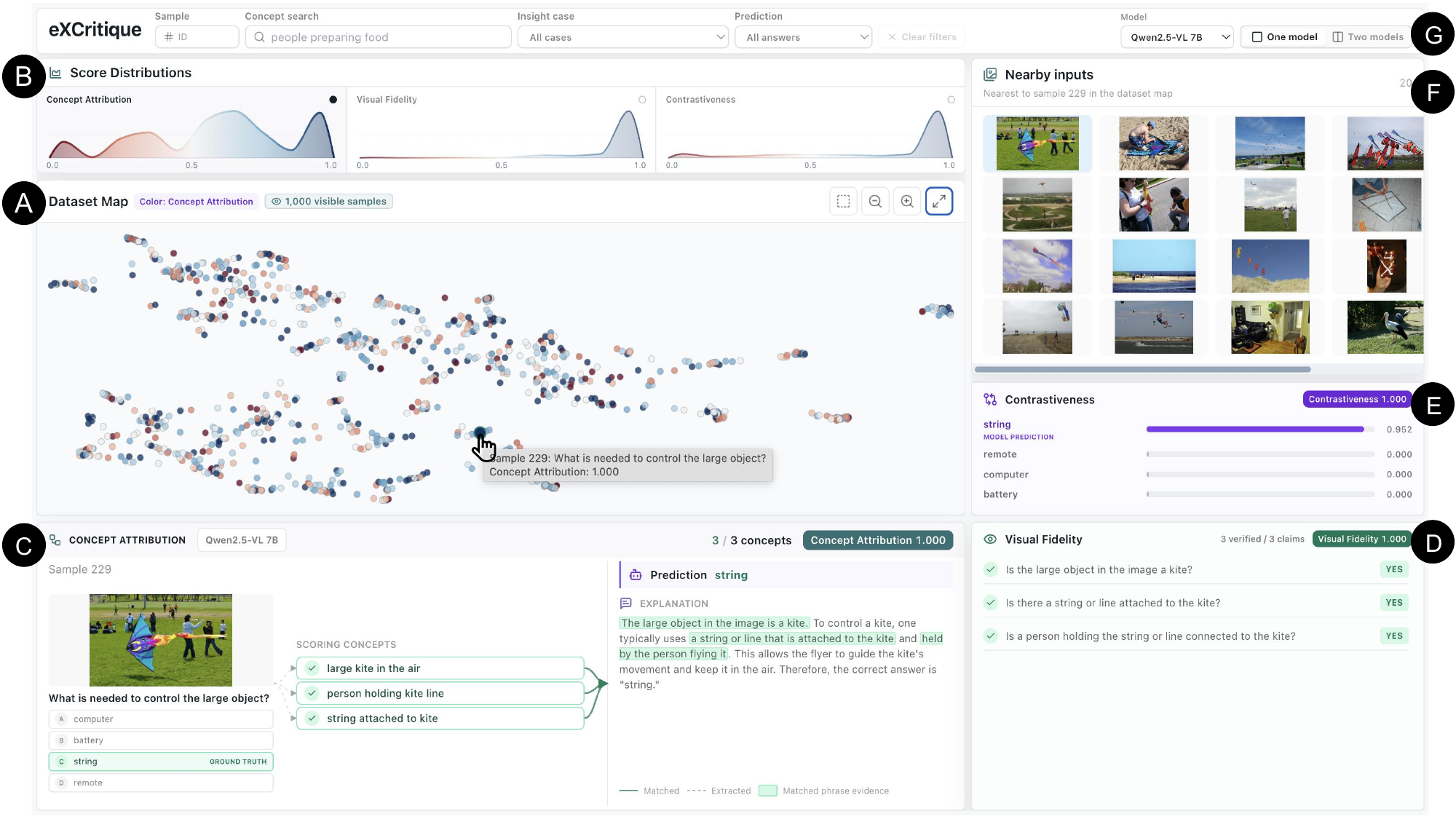}
    \caption{Overview of the \toolname{} interface. (A) Concept-based dataset map; (B) explanation-quality score distributions; (C) Concept Attribution and example details; (D) Visual Fidelity verification results; (E) Contrastiveness scores across answer options; (F) nearby input examples; and (G) search, filtering, and model-selection controls.}
    \label{fig:interface-overview}
\end{figure}

\subsection{Visual Interface}
\label{sec:system-interface}

Figure~\ref{fig:interface-overview} presents an overview of the \toolname{} interface. The \textit{Dataset Map} (A) provides a concept-based representation of the dataset, where each point represents a VQA example and semantically related examples are positioned close to one another. Points can be colored by any of the three explanation-quality metrics, allowing users to observe quality patterns across related examples. Above the map, the \textit{Score Distributions} (B) summarize Concept Attribution, Visual Fidelity, and Contrastiveness scores across the dataset, providing an overview of explanation quality along multiple dimensions (DG2). Users can select score ranges directly from these distributions to filter the examples shown in the map (DG3). Additional controls (G) allow users to search by sample ID or concept, filter by prediction correctness, and explore subsets of interest. Five preset insight filters help users identify examples with specific combinations of explanation-quality scores: \textit{Strong across all scores}, \textit{Weak across all scores}, \textit{Visually faithful, weak contrast}, \textit{Contrastive, visually unsupported}, and \textit{Concept rich, visually unsupported}. 


Selecting an example in the map opens its detailed evaluation results, allowing users to inspect the evidence behind its automated scores (DG4, DG5). The \textit{Concept Attribution} panel (C) presents the input image, question, answer options, model prediction, and generated explanation, along with the extracted visual concepts and their matches in the explanation (DG5). Highlighted phrases help users examine which relevant concepts are covered or omitted. The \textit{Visual Fidelity} panel (D) displays the verification questions derived from the explanation and the corresponding judge-model responses. Similarly, the \textit{Contrastiveness} panel (E) shows how strongly the explanation supports the predicted answer relative to the alternatives. 


\begin{figure}
    \centering
    \includegraphics[width=0.95\linewidth]{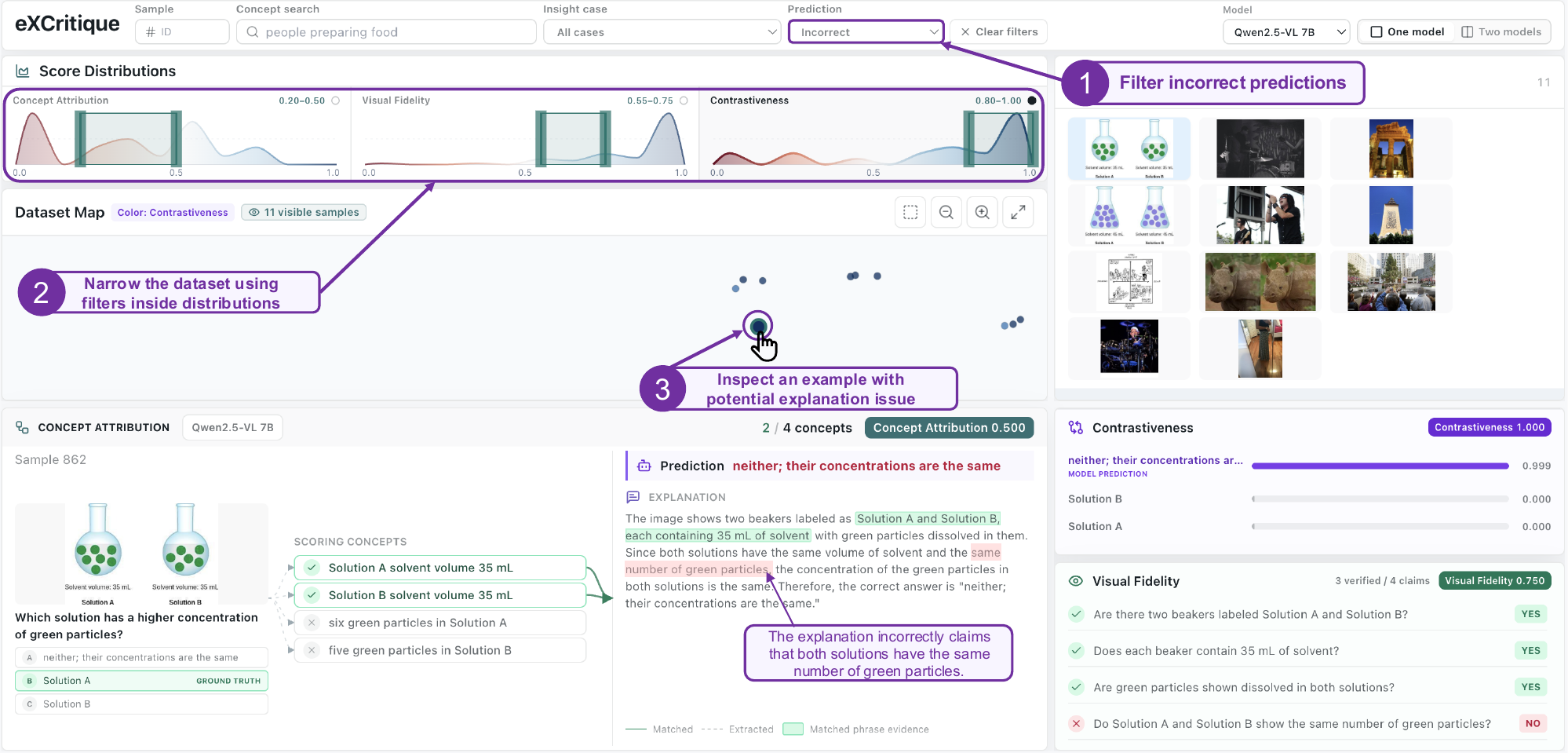}
    \caption{An example of interactive explanation evaluation using \toolname{}. The user (1) filters incorrect predictions, (2) selects ranges across the three quality-score distributions, and (3) inspects an example containing an incorrect explanation. Although the model predicts the wrong answer, the explanation receives Concept Attribution, Visual Fidelity, and Contrastiveness scores of 0.50, 0.75, and 1.00, respectively.}
    \label{fig:interface-interactions}
\end{figure}

Figure~\ref{fig:interface-interactions} illustrates how a user can investigate explanation-quality issues through interactive exploration. First, the user filters the dataset to show only incorrect predictions (Step 1). Next, the user selects score ranges of 0.20--0.50 for Concept Attribution, 0.55--0.75 for Visual Fidelity, and 0.80--1.00 for Contrastiveness (Step 2), narrowing the dataset to 11 examples. From these results, the user selects sample 862 for closer inspection (Step 3). The question asks which of two solutions has a higher concentration of green particles. Although Solution A contains more green particles than Solution B at the same solvent volume, the model incorrectly predicts that their concentrations are equal. The user study reports more illustrative uses of \toolname{}.


\subsection{System Implementation}
\label{sec:system-implementation}

We implemented \toolname{} as a web-based visual analytics system with a React/TypeScript frontend and a Python/FastAPI backend. The frontend uses D3.js to render the visualizations and support the interactions. 
To construct the visual map, we encode  visual concepts from an image using a sentence-transformers model \path{all-MiniLM-L6-v2}. We average and normalize the concept embeddings for each example and project them into two dimensions using UMAP with cosine distance ($n_{\text{neighbors}}=15$, $\text{min\_dist}=0.1$, and random seed $42$).

We generated answers and explanations using the evaluating model, following the two-step procedure of He et al.~\cite{he2026believing}, where the model first predicts an answer and then generates an explanation conditioned on that prediction. We also adopted their prompts and scoring procedures for Visual Fidelity and Contrastiveness. Our Concept Attribution pipeline uses separate prompts to extract question-relevant visual concepts and to match them in the explanation. Further implementation details and prompts are provided in Appendix~\ref{app:prompts}.

\section{User Study}
\label{sec:user-study}

We conducted a within-subject user study to investigate whether interactive exploration of VLM-generated explanations and automated quality scores can support practitioners in evaluating explanation quality. 







\subsection{Study Design}
\label{sec:study-design}

\textbf{Participants.}
We recruited 16 participants (P1--P16; 13 male and 3 female) with experience in developing or evaluating machine learning models and applications. Participants included one undergraduate student, 12 PhD students with relevant research and prior industry experience, two research scientists from industry, and one postdoctoral researcher. The study was approved by the Institutional Review Board (IRB), and informed consent was obtained from all participants before the study.

\noindent
\textbf{Conditions.}
We compared \toolname{} with a spreadsheet-style baseline. Both interfaces provided access to the same types of information, including input images, questions, model predictions, generated explanations, automated quality scores, and supporting evaluation details. The baseline also displayed overall score distributions and supported basic sorting and inspection of individual examples. Unlike \toolname{}, however, it did not provide interactive filtering or coordinated exploration through the visualizations. This setup allowed us to examine the role of interactive exploration while keeping the available evaluation information comparable between conditions. 



\noindent
\textbf{Study Materials.}
We utilized two datasets: A-OKVQA validation split~\cite{schwenk2022aokvqa} and MMStar~\cite{chen2024mmstar}, respectively. For both datasets, we used Qwen2.5-VL-7B to generate answers and explanations. Each example was associated with three precomputed explanation-quality scores: Concept Attribution, Visual Fidelity, and Contrastiveness. The identity of the VLM  and datasets was not disclosed to participants during the study. Each participant evaluated both interfaces using different datasets, with interface order and dataset assignment counterbalanced across participants.

\subsection{Task and Procedure}
\label{sec:study-procedure}

Participants were asked to take the role of an ML developer responsible for evaluating a VLM's explanation quality before deployment. They were informed that the model-generated explanations and automated quality scores had already been computed, and their task was to inspect these results for identifying potential issues.

Participants completed one task using each interface: \emph{identify as many explanation-quality issues or questionable automated scores as possible within ten minutes}. They were encouraged to investigate examples with unusually high or low quality scores, disagreements among the three metrics, or explanations that appeared inconsistent with the available evidence. Participants were asked to think aloud and explain the issues they identified, including the evidence supporting their judgments. 

The study was conducted remotely over Zoom. After providing consent, participants received an introduction to VLM-generated explanations and the three quality metrics through illustrative examples. The moderator then introduced the first assigned condition and demonstrated its main features for approximately five minutes. Participants then acquainted themselves with the assigned condition. They were encouraged to ask questions at this stage. Once comfortable with the interface, participants explored the assigned dataset for ten minutes while thinking aloud. The moderator noted the examples and issues identified, including sample identifiers and timestamps when possible. Participants then completed a questionnaire with likert-scale (scale of 1 to 7) statements to reflect on their experience with the interface. We also administered the System Usability Scale (SUS)~\cite{brooke1996sus} and NASA Task Load Index (NASA-TLX)~\cite{hart1988nasatlx} at this stage.

The same procedure was repeated with the second condition using the other dataset. After completing both conditions, participants took part in a semi-structured interview discussing their preferences, the features they found useful or distracting, and situations in which they would choose one interface over the other. With the consent of the participants, we recorded the sessions, including audio, video, and shared-screen activity.  Participants were compensated at a rate of \$25 per hour, prorated to their session duration. Each session lasted around 60-75 mins.

\subsection{Measures and Data Collection}
\label{sec:study-measures}

For measuring task performance, we recorded two variables. \emph{Unique examples explored} counted distinct examples opened and inspected, including their predictions, explanations, and quality scores. \emph{Participant-identified potential issues} counted distinct examples for which participants explicitly raised concerns about the explanations, automated scores, or their consistency with the available evidence. Repeated inspections or multiple concerns about the same example were counted once. These measures reflect participant judgments rather than independently verified errors.


We compared quantitative variables using two-sided paired $t$-tests, reporting means, standard deviations, paired mean differences with 95\% confidence intervals, and Cohen's $d_z$. The likert-scale statements were analyzed using two-sided Wilcoxon signed-rank tests, with unadjusted exploratory $p$-values and matched-pairs rank-biserial correlations. We also reviewed moderator notes and interview transcripts to identify recurring evaluation strategies, concerns, and interface preferences. During the study, interfaces were identified using neutral labels. For clarity, references to these labels in participant quotations were replaced with the corresponding interface names.
\section{Results}
\label{sec:results}


\subsection{Quantitative Results}
\label{sec:results-rq1}

\begin{figure}[t]
  \centering
  \includegraphics[width=\linewidth]{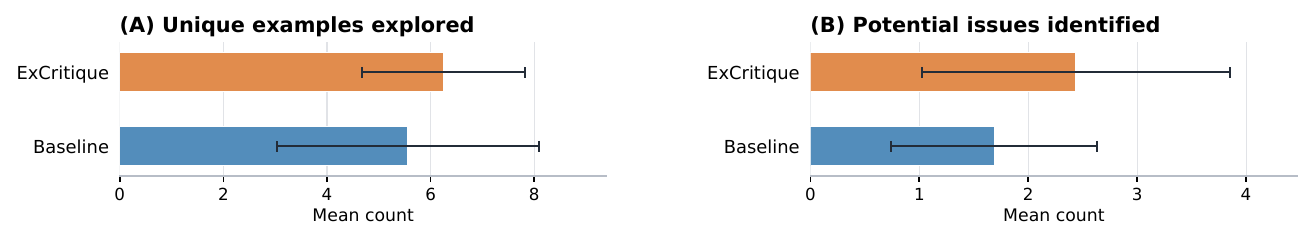}
  \caption{Task performance in the two conditions: (A) number of unique examples explored and (B) number of participant-identified potential issues. Bars show means across 16 participants; error bars show standard deviations.}
  \label{fig:behavioral-results}
\end{figure}

Participants explored an average of 6.25 unique examples ($SD=1.57$) with \toolname{}, compared with 5.56 ($SD=2.53$) with the baseline (Fig.~\ref{fig:behavioral-results}A). This difference was not statistically significant ($\Delta M=0.69$, 95\% CI $[-0.71,\,2.09]$, $p=.312$, $d_z=0.26$). In contrast, participants identified more potential issues with \toolname{} ($M=2.44$, $SD=1.41$) than with the baseline ($M=1.69$, $SD=0.95$; Fig.~\ref{fig:behavioral-results}B). The paired comparison was statistically significant ($\Delta M=0.75$, 95\% CI $[0.04,\,1.47]$, $p=.04$, $d_z=0.56$). 


The results can be explained based on how participants found the examples and patterns. \toolname{} provided clear structures and affordances to identify potential patterns. Thus, while using \toolname{}, participants quickly identified interesting examples and spent more time analyzing them. In contrast, while using the baseline tool, participants jumped from one example to another randomly to identify interesting examples, which ultimately resulted into increased coverage of examples. Thus, the difference between the number of unique examples explored was not statistically significant (Figure~\ref{fig:behavioral-results}A), even though participants flagged more examples as potentially problematic using \toolname{} (Figure~\ref{fig:behavioral-results}B). 
Participants mentioned that they would prefer to use \toolname{} for long-term use and would become more efficient with it over time in finding problematic issues. They had limited time (10 mins) to complete the tasks. Although we provided them with introductory tutorial, there was a learning curve to understand the metrics and interactions in the tool. Thus, the difference we see in Figure~\ref{fig:behavioral-results}B will become larger over time. P05's comment captures this dimension: \textit{``Well, it took me some time to figure out how to find the problematic examples [with \toolname{}], but once I did, it was pretty easy. If I use it daily, I think I should be pretty quick with it.'' } 

\begin{figure*}[t]
    \centering
    \includegraphics[width=0.9\textwidth]{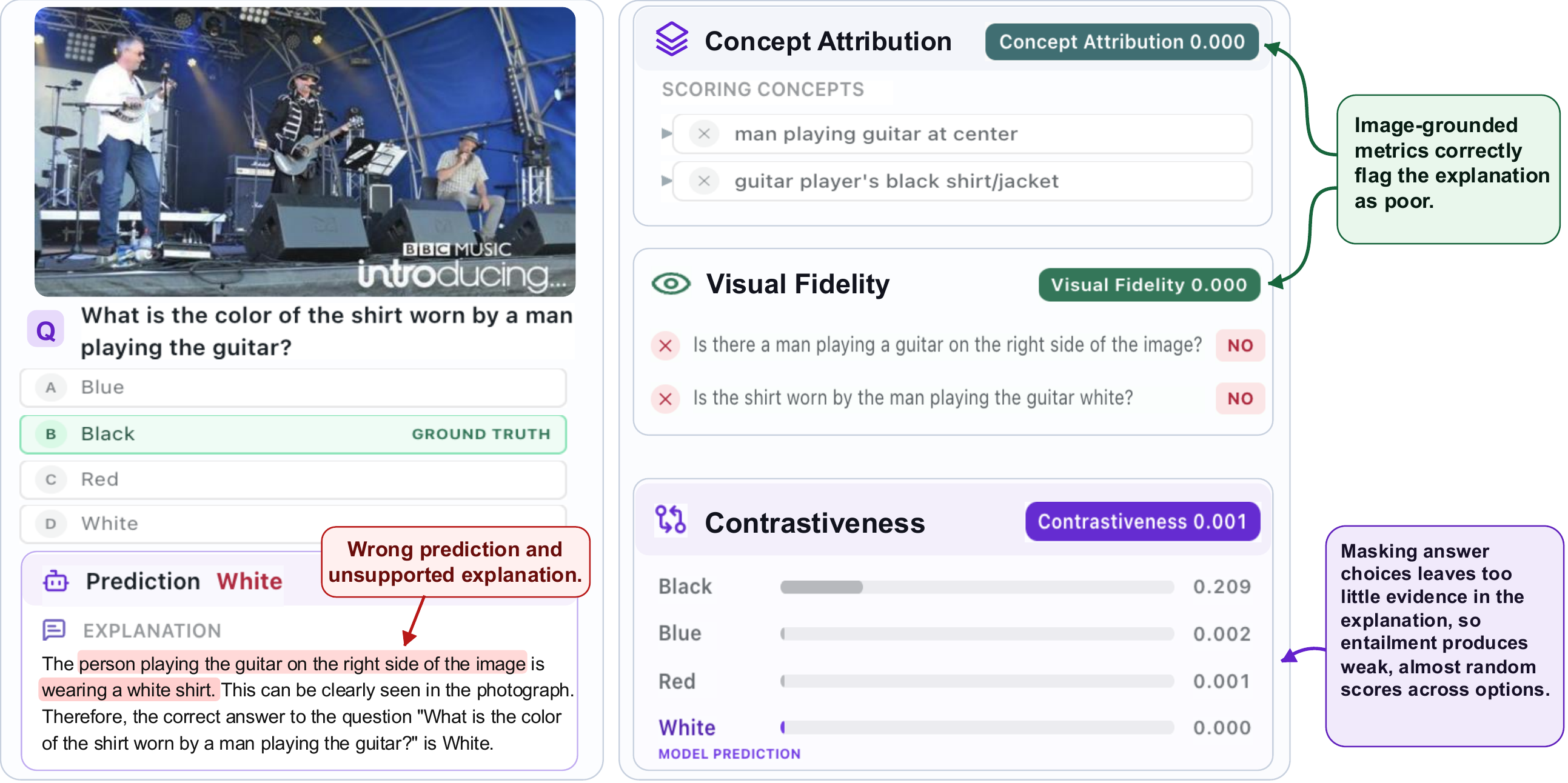}
    \caption{An incorrect explanation with low quality scores and a masking limitation. The model predicts a white shirt instead of the ground-truth black shirt; all three metrics assign near-zero scores. However, masking the answer-color terms leaves little discriminative evidence for the entailment-based Contrastiveness calculation, complicating interpretation of its score.}
    \label{fig:white-shirt-example}
\end{figure*}

\begin{figure*}[t]
    \centering
    \includegraphics[width=0.9\textwidth]{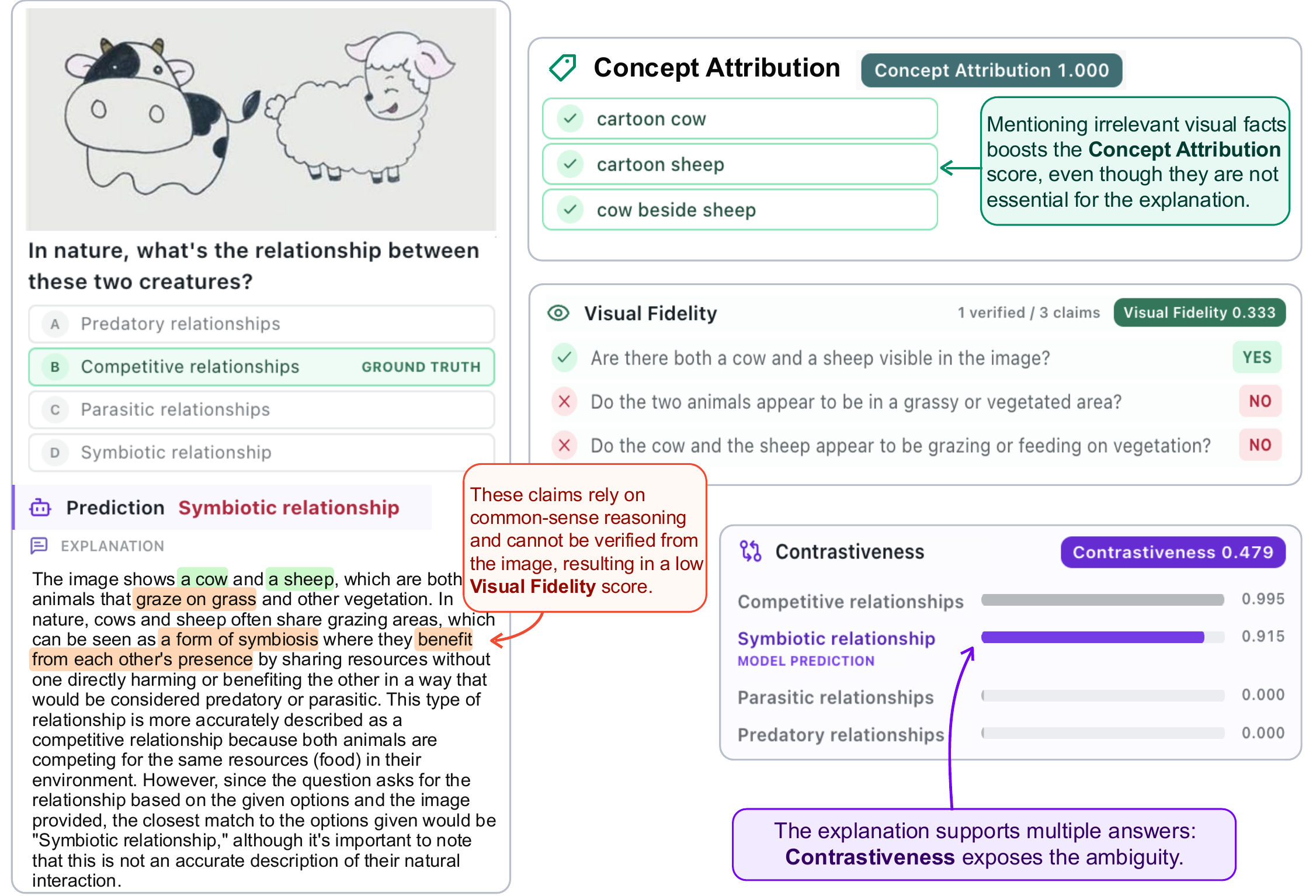}
    \caption{A problematic relationship explanation with differing metric assessments. The model predicts symbiosis rather than the ground-truth competitive relationship. Concept Attribution rewards mentioned visual concepts (1.000), while Visual Fidelity (0.333) flags claims not established by the image and Contrastiveness (0.479) reveals support for competing answers. The example also exposes the limits of verifying commonsense reasoning through image evidence alone.}
    \label{fig:cow-sheep-example}
\end{figure*}
\subsection{Qualitative Feedback}
\label{sec:results-rq2}

\subsubsection{Critiquing Explanations}
Participants approached explanation evaluation in different ways. With the baseline, several relied on sorting the metric columns and opening examples sequentially. With \toolname{}, participants could instead combine prediction-correctness filters, score-range brushing, and preset filters to narrow the dataset before inspecting individual outputs. P08 explained that the most useful feature was the ability to \emph{``filter by multiple criteria,''} particularly when seeking examples in the middle of a score distribution that would be difficult to locate by sorting alone. P15 described a similar contrast when searching for cases with high Contrastiveness and low Concept Attribution: \emph{``I can just do it with two clicks, whereas for the baseline I had to like scroll through it, which was very tedious.''} Participants such as P02, P06, and P13 also experimented with different score ranges or preset filters, illustrating how targeted searches formed part of their evaluation process.

For some participants, however, exploration was less about locating a particular example than understanding the behavior of the metrics across the dataset. P07 selected regions of the map, examined nearby examples, and compared how score distributions changed after filtering. P11 repeatedly combined correct/incorrect prediction filters with score-range selections to investigate relationships among the three metrics. Explaining this interest, P11 said that \emph{``the distribution itself is the most useful thing for me,''} because the filters made it possible to inspect patterns in the scores.


Figures~\ref{fig:white-shirt-example} and~\ref{fig:cow-sheep-example} illustrate the types of explanation problems that such inspection can reveal. In the white-shirt example (Figure~\ref{fig:white-shirt-example}), the model incorrectly predicts white rather than black and provides an explanation with unsupported visual claims. All three quality scores are near zero (Concept Attribution: 0.000; Visual Fidelity: 0.000; Contrastiveness: 0.001), consistent with the poor explanation. In the cow--sheep example (Figure~\ref{fig:cow-sheep-example}), the model incorrectly predicts a symbiotic relationship while its explanation also discusses competition. Visual Fidelity (0.333) and Contrastiveness (0.479) indicate weaknesses, despite a perfect Concept Attribution score (1.000). Together, the examples show how scores and instance-level evidence can be examined jointly when critiquing model explanations.

\subsubsection{Critiquing Automated Scores}
The ability to inspect individual score calculations prompted participants to scrutinize the automated scores themselves. Several questioned Concept Attribution when extracted concepts appeared irrelevant to the question or when relevant ideas were expressed in different words. P08 observed that \emph{``The key concepts are a bit Vague,''} and questioned the importance of some concepts selected by the judge. P06 suggested weighting concepts by their relevance rather than treating all of them equally. P12 identified another limitation: for questions requiring commonsense reasoning, a useful explanation may contain information that is not directly visible in the image, even though the concept list is extracted from visual evidence.

The illustrative example in Figure~\ref{fig:cow-sheep-example} highlights these limitations. Concept Attribution assigns full credit for identifying the cow and sheep, even though these concepts do not establish the relationship in question. Visual Fidelity exhibits a related limitation, as claims about grazing and animal relationships rely on contextual or commonsense knowledge that cannot be verified from the image alone, resulting in a low score (0.333). Finally, Contrastiveness assigns high entailment scores to both competitive (0.995) and symbiotic (0.915) relationships, indicating that the explanation does not clearly distinguish between the alternatives.

Participants also raised concerns about the verification steps underlying Visual Fidelity and Contrastiveness. P01, P05, and P08 noted instances where Visual Fidelity generated questions they considered uninformative or irrelevant; P12 questioned some verifier responses to image-based questions. In their review of Contrastiveness, P08, P11, P12, P13, and P16 encountered cases where masking answer terms appeared to leave little evidence for the entailment model to evaluate. P15 summarized the need for refinement by saying that \emph{``concept attribution and visual fidelity are good metrics, but they have room for improvements''} in how concepts are selected and questions are verified.

A related limitation is observed in Figure~\ref{fig:white-shirt-example}, where a low metric score highlights potential issues but requires careful interpretation. Concept Attribution and Visual Fidelity appropriately identify unsupported visual claims. However, masking the answer choices removes the primary evidence used by the explanation to distinguish between shirt colors. So, the low Contrastiveness score (0.001) is difficult to interpret without considering the masking procedure and should not be treated as a definitive indicator of explanation quality.



\subsection{Perceived Usefulness, Usability, and Workload}
\label{sec:results-rq3}

\begin{figure}[t]
  \centering
  \includegraphics[width=0.9\linewidth]{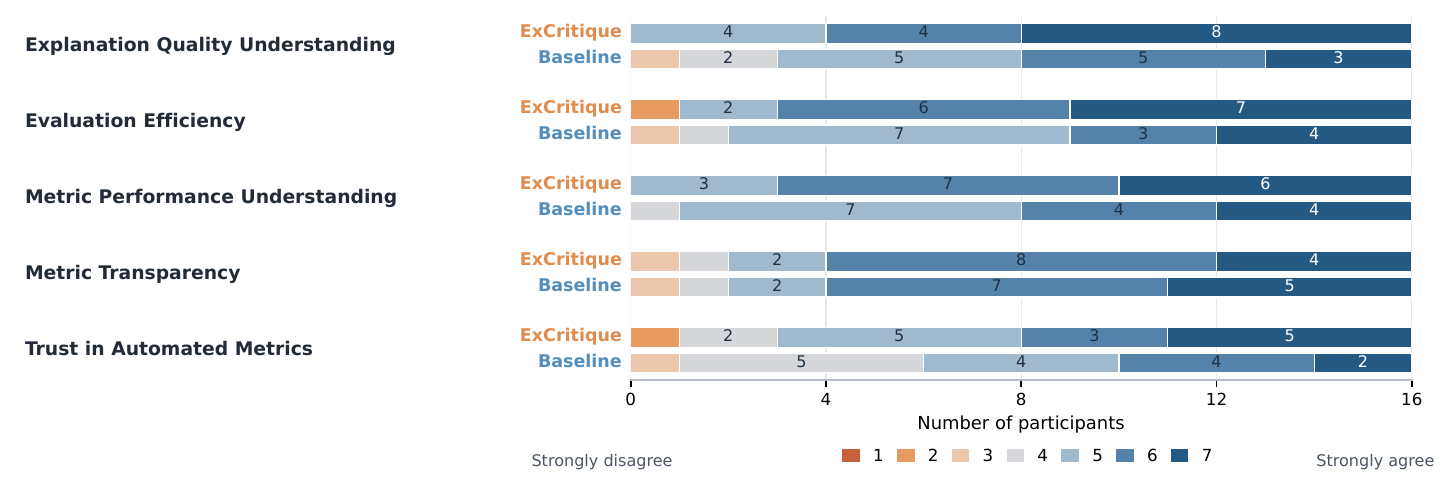}
  \caption{Distributions of the five study-specific seven-point Likert ratings for \toolname{} and the baseline ($N=16$ per condition). Each segment represents the number of participants selecting that response. Wilcoxon comparisons are exploratory and unadjusted.}
  \label{fig:likert-results}
\end{figure}

Figure~\ref{fig:likert-results} shows the distributions of responses to the five study-specific questions. Participants rated \toolname{} more highly for understanding explanation quality (median $6.5$ versus $5.5$; Wilcoxon $p=.015$, rank-biserial $r_{rb}=0.71$) and understanding the performance of automated metrics (median $6.0$ versus $5.5$; $p=.011$, $r_{rb}=1.00$). Differences for evaluation efficiency ($p=.150$, $r_{rb}=0.53$), metric transparency ($p=.658$, $r_{rb}=-0.14$), and trust in automated metrics ($p=.254$, $r_{rb}=0.40$) were not statistically significant. 

Descriptive scores for usability and workload were similar between conditions. 
The baseline had a slightly higher SUS mean ($M=72.50$, $SD=10.95$) than \toolname{} ($M=69.69$, $SD=12.91$), whereas raw NASA-TLX scores were nearly identical ($M=27.76$, $SD=16.74$ for the baseline; $M=26.98$, $SD=18.10$ for \toolname{}).

We noticed several usability issues with \toolname{} in the study. As mentioned earlier, \toolname{} had a learning curve, especially for users who were not familiar with the quality metrics. Some participants mentioned that the interactive map was a bit cluttered and that it was difficult to pinpoint specific data points with hover actions in some cases. They suggested decluttering the map and increasing dot sizes. P07 and P11 suggested resizing the panels, and P05 requested more precise numerical controls for selecting score ranges. P16 also asked for clearer guidance on interpreting metric values, such as examples illustrating what scores of $0.5$ or near zero mean.

\section{Discussion}

\toolname{} opens up research in several directions. Future research can explore critiquing explanations for a broader set of tasks. We explored critiquing explanations for a specific type of task (visual question answering, or VQA). The quality metrics were also specific to the task. This design choice helped us validate the idea of visual interfaces for facilitating human-AI collaborative critiquing of explanations. Future research can explore extending \toolname{} to other tasks such as text summarization, meta-data extraction, video surveillance, and so on. Each of these tasks would require custom quality metrics that are beyond the scope of this paper. Through exploration of the tasks, we can move towards a unified and generalized visual interface for critiquing explanations generated by AI Models for any tasks.

Another way to extend \toolname{} is to think about developing mechanisms for human oversight over automated judge approaches. Both academia and industry are increasingly using automated judge approaches for LLM-based tasks. While this practice has accelerated research and technology outputs, it has also introduced a new task: \textit{judging} the \textit{judge}~\cite{DBLP:conf/nips/ZhengC00WZL0LXZ23, croxford2025evaluating}. Tools like \toolname{} can be transformative in how researchers and practitioners evaluate automated judge approaches and metrics.

Our work opens up several directions for HCI and FAccT research too. Our results suggest that \toolname{} improved several normative aspects of human-AI collaboration (e.g., model understanding). However, the study setting was controlled and did not allow users to extensively use \toolname{}. Future research can design more studies to measure these normative aspects systematically. Previous research on measuring trust, ownership, and agency could motivate this thread of research~\cite{10.1145/3313831.3376624, DBLP:conf/chi/HoqueMGSCKE24}.
\section{Conclusion}
We presented \toolname{}, an interactive tool for critiquing AI-generated explanations. The tool provides visual representations and interactions to help researchers and practitioners simultaneously identify issues with explanations and quality metrics. We validated the tool in a controlled study where we compared it with a baseline tool. The findings suggest that participants were able to identify more problematic examples using \toolname{}, compared to the baseline, even within a short, time-constrained task. Participants were excited about using tools such as \toolname{} for long-term use in practice and mentioned that these tools will significantly improve their workflow and efficiency. Overall, this work provides a promising approach for validating open-ended text generation and automated quality metrics.
\section{Endmatter}

\subsection{Generative AI Disclosure Statement}

We used OpenAI Codex (GPT-5.6 Sol) to assist with implementing the \toolname{} interface, following author-defined designs and specifications. The authors guided the development process and manually reviewed the generated code. ChatGPT (GPT-6) was used to identify relevant literature and all suggested references were independently reviewed and verified by the authors. GPT-6 was also used to edit and refine language and grammar in selected sections of the manuscript, with all revisions reviewed by the authors. Additionally, GPT-6 assisted in writing Python code for statistical analysis and visualizations. The authors reviewed and adapted the code and executed the analyses locally without sharing study data with the AI tool. The authors retain full responsibility for the research, results, and manuscript.

\subsection{Ethical Considerations Statement}

This study received approval from an institutional ethics review board before data collection. Participants provided verbal informed consent and participated voluntarily, with the option to skip questions, take breaks, or withdraw without penalty. Audio and screen recordings were collected only with permission. To protect privacy, we used coded participant identifiers and stored recordings, transcripts, and survey data in secure institutional storage accessible only to authorized researchers. No participant data were shared with the AI models used in the study.

The study posed minimal risk, primarily involving possible fatigue or discomfort during evaluation, which participants could manage by pausing or withdrawing. Beyond the study, automated explanation-quality scores may create misplaced confidence if treated as definitive assessments. To mitigate this risk, \toolname{} presents scores alongside the underlying explanations and supporting evidence, allowing practitioners to inspect and question automated evaluations rather than accept them as ground truth.

\bibliographystyle{ACM-Reference-Format}
\bibliography{references}


\appendix

\section{Model Configurations and Prompts}
\label{app:prompts}

\subsection{Model Configurations}
\label{app:model-configurations}

We used Qwen2.5-VL-7B, accessed through Hugging Face, to generate answers and explanations for multiple-choice VQA examples. The maximum generation length was 20 new tokens for answers and 4096 new tokens for explanations, with temperature $0.1$ for both stages.

For LLM-based quality evaluation, we used \path{gpt-5.4-2026-03-05} through the OpenAI API, with temperature $0.1$ and a maximum output length of 1024 tokens. This model replaced the GPT-4o and GPT-4o-mini models used in the corresponding judging stages of He et al.~\cite{he2026believing}. Contrastiveness retained the entailment verifier \path{soumyasanyal/nli-entailment-verifier-xxl} and the \path{google/flan-t5-xxl} tokenizer, accessed through Hugging Face. We also used the same GPT-5.4 model for Concept Attribution's concept extraction and semantic matching.

\subsection{Answer and Explanation Generation}
\label{app:generation-prompts}

Following He et al.~\cite{he2026believing}, we generated post-hoc explanations in two steps. First, the VLM predicted an answer from the image, question, and answer choices. It then generated an explanation conditioned on that answer. We used the prompts reported in their Appendix C, Table 3.

\subsection{Visual Fidelity and Contrastiveness}
\label{app:reused-prompts}

We adopted the prompting and scoring procedures of He et al.~\cite{he2026believing} without changing their prompt wording. Visual Fidelity generates verification questions from the explanation and checks each question against the image; the corresponding prompts appear in their Tables 7 and 8. Contrastiveness masks answer mentions, converts the question and each answer choice into declarative hypotheses, and compares their entailment probabilities using the verifier specified above. The question--answer reformulation prompt is reported in their Table 5. The changes to the LLM judge models are described in Appendix~\ref{app:model-configurations}.

\subsection{Concept Attribution Prompts}
\label{app:ca-prompts}

Concept Attribution uses two prompts. The first extracts up to seven directly observable concepts relevant to the image, question, and answer choices. The second checks whether each concept is supported by the generated explanation, accepting explicit mentions and semantic paraphrases but rejecting contradictions or merely related wording. For matched concepts, the evaluator returns a binary decision and the shortest exact supporting passage. Tables~\ref{tab:ca-extraction} and~\ref{tab:ca-matching} provide the complete prompts.

\begingroup
\small
\setlength{\LTpre}{6pt}
\setlength{\LTpost}{8pt}
\begin{longtable}{@{}>{\raggedright\arraybackslash}p{\linewidth}@{}}
\caption{Concept Extraction: System and User prompts.}\label{tab:ca-extraction}\\
\toprule
\textbf{System prompt} \\[2pt]
You are an expert visual reasoning assistant. \\[6pt]
\midrule
\textbf{User prompt} \\[4pt]
Examine the image carefully with the VQA question and answer choices. \\[5pt]
Extract visible facts that directly help answer the question. Focus on the target object, property, or relationship. Include surrounding context only when it helps distinguish the possible answers. \\[5pt]
Write each concept as a short, specific phrase, usually 2-7 words, rather than a full sentence. Each phrase should capture one directly observable object, attribute, action, count, readable text, or spatial relation. Retain the object and essential relation so the phrase is unambiguous (e.g. "fork tines pointing toward broccoli"). \\[5pt]
Before returning, remove unrelated details, duplicate evidence, combined facts, and choice-elimination arguments. Do not guess unclear properties, or infer visual facts from the answer choices. There is no minimum number of concepts; include only those that are relevant and important for answering the question. Return only JSON with a "concepts" list. \\[5pt]
\textbf{Example 1:} \\[3pt]
\textbf{Question:} What color is the umbrella held by the woman? \\[2pt]
\textbf{Choices:} red, blue, yellow, green \\[2pt]
\textbf{Image:} A woman holds a blue umbrella beside trees and pedestrians. \\[2pt]
\textbf{Output:} \{\{"concepts": ["woman's blue umbrella"]\}\} \\[2pt]
\textbf{Example 2:} \\[3pt]
\textbf{Question:} Which flag is displayed on the wall? \\[2pt]
\textbf{Choices:} American, French, English, Moravian \\[2pt]
\textbf{Image:} A flag with red and white stripes and white stars on a blue field hangs beside a clock. \\[2pt]
\textbf{Output:} \{\{"concepts": ["red-white flag stripes", "white stars on blue flag field"]\}\} \\[2pt]
\textbf{Question:} \{question\} \\[2pt]
\textbf{Choices:} \{choices\} \\[2pt]
\bottomrule
\end{longtable}
\endgroup

\begingroup
\small
\setlength{\LTpre}{6pt}
\setlength{\LTpost}{8pt}
\begin{longtable}{@{}>{\raggedright\arraybackslash}p{\linewidth}@{}}
\caption{Semantic Matching: System and User prompts.}\label{tab:ca-matching}\\
\toprule
\textbf{System prompt} \\[2pt]
You are a careful semantic attribution evaluator. \\[6pt]
\midrule
\textbf{User prompt} \\[4pt]
For each concept, decide whether a word, phrase, or sentence in the explanation supports it. Accept explicit descriptions, synonyms, paraphrases, and clear contextual implications. Identical wording is unnecessary, but the support must describe the same object and visual fact. \\[5pt]
A related word alone, a statement about another object, or a contradiction does not establish a match. Repeating the answer supports a concept only when it also expresses that concept's visual fact. \\[5pt]
For each present concept, copy the shortest exact supporting passage into evidence\_quote, keeping enough context to identify its referent. Copy the explanation's wording rather than rewriting it. For an absent or contradicted concept, set present=false and evidence\_quote="". \\[5pt]
Return only JSON with a "matches" list containing one entry per concept in the original order. Each entry must contain concept, present, and evidence\_quote. \\[5pt]
\textbf{Example 1:} \\[3pt]
\textbf{Concepts:} ["left bin's reflective surface", "left bin's vertical ribs"] \\[2pt]
\textbf{Explanation:} "The container on the left catches the light, suggesting steel." \\[2pt]
\textbf{Output:} \{\{"matches": [\{\{"concept": "left bin's reflective surface", "present": true, "evidence\_quote": "The container on the left catches the light"\}\}, \{\{"concept": "left bin's vertical ribs", "present": false, "evidence\_quote": ""\}\}]\}\} \\[2pt]
\textbf{Example 2:} \\[3pt]
\textbf{Concepts:} ["cyclist's protective headwear", "wet pavement"] \\[2pt]
\textbf{Explanation:} "The cyclist is wearing a helmet while riding on a dry street." \\[2pt]
\textbf{Output:} \{\{"matches": [\{\{"concept": "cyclist's protective headwear", "present": true, "evidence\_quote": "The cyclist is wearing a helmet"\}\}, \{\{"concept": "wet pavement", "present": false, "evidence\_quote": ""\}\}]\}\} \\[2pt]
\textbf{Concepts:} \{concepts\} \\[2pt]
\textbf{Explanation:} \{explanation\} \\[2pt]
\bottomrule
\end{longtable}
\endgroup

\end{document}